\documentclass{aastex62}

\received{March 26, 2018}
\revised{April 22, 2018}
\accepted{April 23, 2018}
\submitjournal{ApJ}

\shorttitle{Compound Solar Eruption}
\shortauthors{Dhakal et al.}
\begin{document}

\title{\textbf{A Study of a  Compound Solar Eruption with Two Consecutive Erupting Magnetic Structures}}

\correspondingauthor{Suman Dhakal}
\email{sdhakal2@gmu.edu}

\author{Suman K. Dhakal}
\affiliation{Department of Physics and Astronomy, George Mason University, 4400 University Dr., MSN 3F3, Fairfax, VA 22030, USA}

\author{Georgios Chintzoglou}
\affiliation{Lockheed Martin Solar and Astrophysics Laboratory, Palo Alto, CA 94304, USA}
\affiliation{University Corporation for Atmospheric Research, Boulder, CO 80307-3000, USA}

\author{Jie Zhang}
\affiliation{Department of Physics and Astronomy, George Mason University, 4400 University Dr., MSN 3F3, Fairfax, VA 22030, USA}

\begin{abstract}
We report a study of a compound solar eruption that was associated with two consecutively erupting magnetic structures and correspondingly two distinct peaks, during impulsive phase, of an M-class flare (M8.5). Simultaneous multi-viewpoint observations from $\textit{SDO}$, $\textit{GOES}$ and $\textit{STEREO-A}$ show that this compound eruption originated from two pre-existing sigmoidal magnetic structures lying along the same polarity inversion line. Observations of the associated pre-existing filaments further show that these magnetic structures are lying one on top of the other, separated by 12 Mm in height, in a so-called ``double-decker" configuration. The high-lying magnetic structure became unstable and erupted first, appearing as an expanding hot channel seen at extreme ultraviolet wavelengths. About 12 minutes later, the low-lying structure also started to erupt and moved at an even faster speed compared to the high-lying one. As a result, the two erupting structures interacted and merged with each other, appearing as a single coronal mass ejection in the outer corona. We find that the double-decker configuration is likely caused by the persistent shearing motion and flux cancellation along the source active region's strong-gradient polarity inversion line. The successive destabilization of these two separate but closely spaced magnetic structures, possibly in the form of magnetic flux ropes, led to a compound solar eruption. The study of the compound eruption provides a unique opportunity to reveal the formation process, initiation, and evolution of complex eruptive structures in solar active regions. 
\end{abstract}


\section{Introduction} \label{sec:intro}
Solar eruptions, manifested as flares and coronal mass ejections (CMEs), are the most magnificent events happening on the solar atmosphere. They are the main sources of disturbances in interplanetary space and the driver of space weather near the Earth. Flares are sudden and localized emission of radiation over the entire electromagnetic spectrum. They are routinely identified and classified based on their intensity profile of soft X-ray (SXR) emission (\citealt{Fletcher_etal_2011}). CMEs, which have wider effects, are the eruption of magnetized plasma from the solar atmosphere identified as arc-like brightenings in coronagraph images in the outer corona. Flares and CMEs, often occurring together, are manifested as a single energy release process, during which a tremendous amount of stored magnetic energy is released from the solar atmosphere \citep{Forbes_etal_2000}. The general model of eruptive flares (flares with CMEs) describes the solar eruption as the eruption of a magnetic flux rope (MFR). An MFR is a coherent structure of magnetic field twisting around a central axis that carries a large amount of free magnetic energy. A typical solar eruption has three phases in the X-ray profiles: (1) the precursor phase, (2) the impulsive phase (IP), and (3) the gradual phase (see, \citealt{Zhou_etal_2016} for a discussion). This three-phase evolution scenario has been also clearly observed in CMEs and is associated with three distinct acceleration phases of them (e.g., \citealt{Zhang_etal_2001}).

It has been suggested that an MFR can exist before the eruption~(\citealt{van_Ballegooijen_etal_1989};~\citealt{Chen_etal_1997})  or it could form during it (\citealt{Antiochos_etal_1999};~\citealt{Lynch_etal_2008}). If an MFR exists before the eruption, the question is how and where it could form and how to detect it. It has been suggested that an MFR could form by different mechanisms prior to its eruption, such as shearing motions (\citealt{Amari_etal_2000,Amari_etal_I_2003, Jacobs_etal_2009}), flux cancellation at the polarity inversion line (PIL) (\citealt{van_Ballegooijen_etal_1989, Aulanier_etal_2010, Green_etal_2011}), flux emergence (\citealt{Fan_and_Gibson_2003, Leake_etal_I_2013}), and confined flaring (\citealt{Patsourakos_etal_2013}, \citealt{Chintzoglou_etal_2015}). The detection of an MFR is difficult as it has low plasma density, and also due to the technological inability to measure magnetic field directly in the corona. However, there is indirect evidence that may suggest the existence of MFRs in the form of a cavity and prominence (associated with filaments when seen on the solar disk), sigmoidal bright loops in X-rays (\citealt{Rust_and_Kumar_1996}; \citealt{Green_etal_2007}), AIA/$\textit{SDO}$ hot channel structures (\citealt{Cheng_etal_2011}; \citealt{Zhang_etal_2012}), dips or bald patches in filament channels (\citealt{Lites_etal_2005}; \citealt{Lopez_Ariste_etal_2006}), and nonlinear force-free field (NLFFF) extrapolation results obtained for the photospheric boundary conditions (e.g., \citealt{Chintzoglou_etal_2015}). 

The pre-erupting structures (MFRs) are kept in equilibrium by a downward force due to the pressure and magnetic tension from the overlying magnetic field \citep{Chen_1989}.
The instability of MFRs could result in three scenarios: (i) failed eruption, (ii) partial eruption, and (iii) full eruption of the MFR. In a failed eruption, the MFR rises up as it is erupting and then stops at a certain maximum height \citep{Song_etal_2014}. A failed eruption could be due to many factors such as strong overlying field pressure \citep{Sun_etal_2015}, an asymmetric overlying coronal field above the filament \citep{Liu_etal_2009}, or not having enough energy for a successful eruption \citep{Shen_etal_2011}. In partial eruptions, (a) reconnection within the MFR \citep{Gibson_and_Fan_2006} separates it into two parts, or (b) the partial emergence and reconnection with the coronal dipole field may result in the split of the MFR into two parts  \citep{Leake_etal_II_2014}. In partial eruptions, the upper part of the MFR erupts and its lower part remains in place. Non-erupting filaments and the immediate re-formation of long-lived X-ray sigmoids after eruption were considered to be partial eruptions (\citealt{Tang_1986}; \citealt{Gilbert_etal_2000}; \citealt{Gibson_etal_2002}). The full eruption is the case where the MFR fully escapes from the Sun, leaving behind a post-flare loop arcade in a potential magnetic configuration.

In some instances, multiple solar eruptions occur within a short period of time; these are known as sympathetic eruptions. They may occur at different active regions (ARs) or within the same AR with multiple PILs. It is debatable whether these consecutive eruptions are related to each other \citep{Biesecker_and_Thompson_2000}. Interrelation among sympathetic eruptions has been shown in the simulation work by~\citet{Torok_etal_2011}. In their 3D magnetohydrodynamic (MHD) simulation, the eruption of one filament can weaken the overlying magnetic pressure sufficiently, leading to the eruption of the other filaments. There is a high chance of interaction among CMEs when they occur in close proximity and/or within a short period of time. The study by \citet{Gopalswamy_etal_2002} found that the interaction among CMEs could be an important factor for solar energetic particle (SEP) production. For a complex AR with multiple PILs, MFRs could be formed along the different PILs and they could erupt consecutively (e.g., \citealt{Chintzoglou_etal_2015}). There are also cases where, instead of having a single MFR along the PIL, two MFRs could be formed along the same PIL (e.g. \citealt{Liu_etal_2012}). The two MFRs arrange themselves in a double-decker (DD) configuration: one lying above the other along the same PIL. Unlike a partial eruption, in a DD configuration the lower and upper structures are observed to exist well before the commencement of the eruption. The two branches of the DD could interact with each other (e.g. \citealt{Liu_etal_2012}; \citealt{Zhu_etal_2015} ).

Besides sympathetic eruptions, another type of inter-related eruption is known as a homologous event.  CMEs originating from the same region, associated with homologous flares, having similar coronographic appearance and similar dimming in extreme ultraviolet imaging are known as homologous CMEs \citep{Zhang_and_Wang_2002}. Their study may also provide information on the physical mechanism of free-energy build-up and the triggering mechanisms of CMEs (\citealt{Nitta_and_Hudson_2001}; \citealt{Zhang_and_Wang_2002}; \citealt{Vemareddy_2017}). 

In this paper, we analyze solar eruption SOL2012-03-10T16:50:00L303C17, which is associated with a single flare (M8.5) and a single CME. However, the flare contains two peaks in its IP separated by only 12 minutes, and the eruption contains two different erupting structures sharing the same PIL. We name this type of eruption  a ``compound eruption,'' i.e. two closely connected magnetic structures erupting consecutively within a short time interval, forming a single flare and a single CME. The compound eruption studied in this paper originated from NOAA AR 11429 on 2012 March 10. Study of the photospheric magnetic structure showed that the shearing and converging motions led to flux cancellation and new connectivity in the corona. These processes resulted in the formation of two MFR candidates, which were arranged in a DD configuration. The paper is structured as follows. The instruments and data are described in Section~\ref{instru}. The compound eruption is described in Section~\ref{eruption}. The evolution of the AR prior to the eruption and the pre-eruptive configuration are analyzed in Section~\ref{obs}, and the discussion and conclusion are presented in Section~\ref{DC}.

\section{\textbf{Instruments}}
\label{instru}
The data used in this study are primarily from the Atmospheric Imaging Assembly (AIA; \citealt{Lemen_etal_2012}) and Helioseismic and Magnetic Imager (HMI; \citealt{Schou_etal_2012}), onboard the $\textit{Solar Dynamics Observatory}$ ($\textit{SDO}$;~\citealt{Pesnell_etal_2012}). The AIA provides full disk images of the Sun with high cadence (12\,s) and high spatial resolution $1\farcs2$. The 10 passbands of the AIA, covering the temperature range from 60,000 K to 20 MK, observe the Sun from the photosphere, chromosphere and corona up to 1.3 $R_{\sun}$. The HMI provides line-of-sight magnetic field observations at a cadence of 45\,s and full vector magnetic data at a cadence of 720\,s. We used observations of the AR 11429 from 2012 March 9,  04:00 UT to 2012 March 10, 19:00 UT. There were brief moments when the Earth obscured ($\textit{SDO}$ eclipse) the view of the $\textit{SDO}$ on each day, therefore data were absent around that time period. We also used imaging from the Extreme Ultraviolet Imager (EUVI; \citet{Wuelser_etal_2004}) and white light images of the Sun-Earth Connection Coronal and Heliospheric Investigation (SECCHI; \citealt{Howard_etal_2008}) onboard the $\textit{Solar Terrestrial Relations Observatory}$ ($\textit{STEREO}$; \citet{Kaiser_etal_2008}). The $\textit{STEREO}$ mission has two satellites; one moves ahead ($\textit{STEREO-A}$) and the other behind ($\textit{STEREO-B}$) the Earth. During the observational period, AR 11429 was behind the limb for $\textit{STEREO-B}$ and could be seen through $\textit{STEREO-A}$ and the $\textit{SDO}$. In addition, we used observations from the Large Angle Spectrometric Coronagraph (LASCO; \citealt{Brueckner_etal_1995}) on board the $\textit{Solar and Heliospheric Observatory}$ ($\textit{SOHO}$), and the Solar X-ray Imager (SXI;~\citealt{Hill_etal_2005} and \citealt{Pizzo_etal_2005}) on board the $\textit{Geostationary Operational Environmental Satellite}$ ($\textit{GOES}$).

\section{\textbf{Observations of the Compound Eruption}}
\label{eruption}
\begin{figure}[h]
\centering
\includegraphics[width=\linewidth,clip]{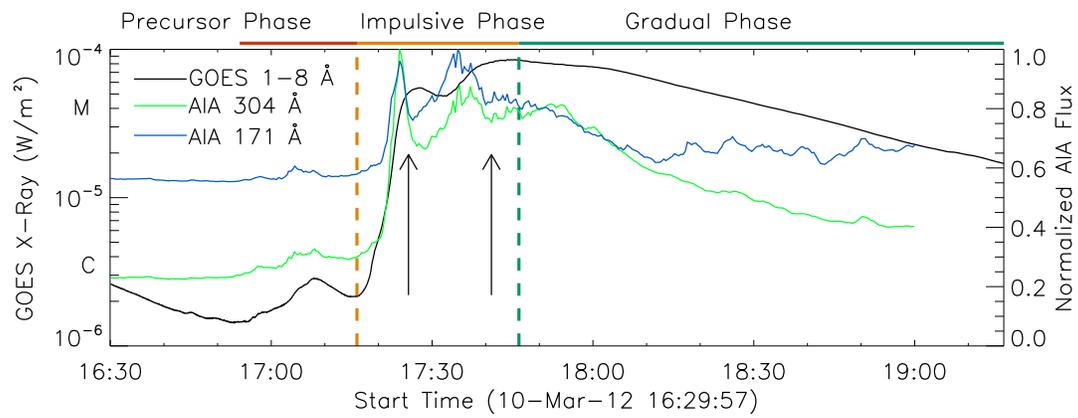}
\caption{Three phases of the compound eruption. The black curve shows the SXR time-profile observed by $\textit{GOES}$. The colored lines show the intensity plot for different AIA passbands. The vertical-dashed lines indicate the beginning and ending of the impulsive phase. The two peaks in the impulsive phase are  indicated by two black arrows.}
\label{fig:three_phase}
\end{figure}

NOAA AR 11429 was a mature AR when the eruption under study occurred. It was located near the center of the solar disk (N17E01) and had a very complex magnetic configuration ($\beta/\gamma/\delta$). The compound eruption began on 2012 March 10, around 16:54 UT, in the southwest (SW) part of the AR. The X-ray profile of the eruption had three distinct peaks; all three were also shown in the AIA pass-bands (see Figure~\ref{fig:three_phase}). The flare could be divided into three phases: precursor phase, IP, and gradual phase. However, unlike typical flares (of which time profiles are single peaked), this flare exhibited two peaks during the IP (indicated by two black arrows in Figure~\ref{fig:three_phase}). AR 11429 had two pre-eruptive hot-channel structures and corresponding filaments on the SW side of the AR. These structures are called hot-channel structures as they are observed in hot passbands of the AIA i.e., 94~\AA{} (6 MK) and 131~\AA{} (10 MK) (e.g., \citealt{Zhang_etal_2012}). These two hot-channel structures (and associated filaments) were lying one above the other along the same PIL in a DD configuration (a detailed study of the  pre-eruptive configuration is in Section~\ref{coronal_configuration}).
Hereafter, the higher-lying hot-channel structure will be referred to as ``hot-channel structure A" (HCSA), and the lower-lying one as ``hot-channel structure B" (HCSB). We also refer to the higher-lying filament as ``filament-A" and lower-lying filament as ``filament-B".

\subsection{\textbf{The Precursor Phase}}
\begin{figure}[h]
\centering
\includegraphics[width=\linewidth,clip]{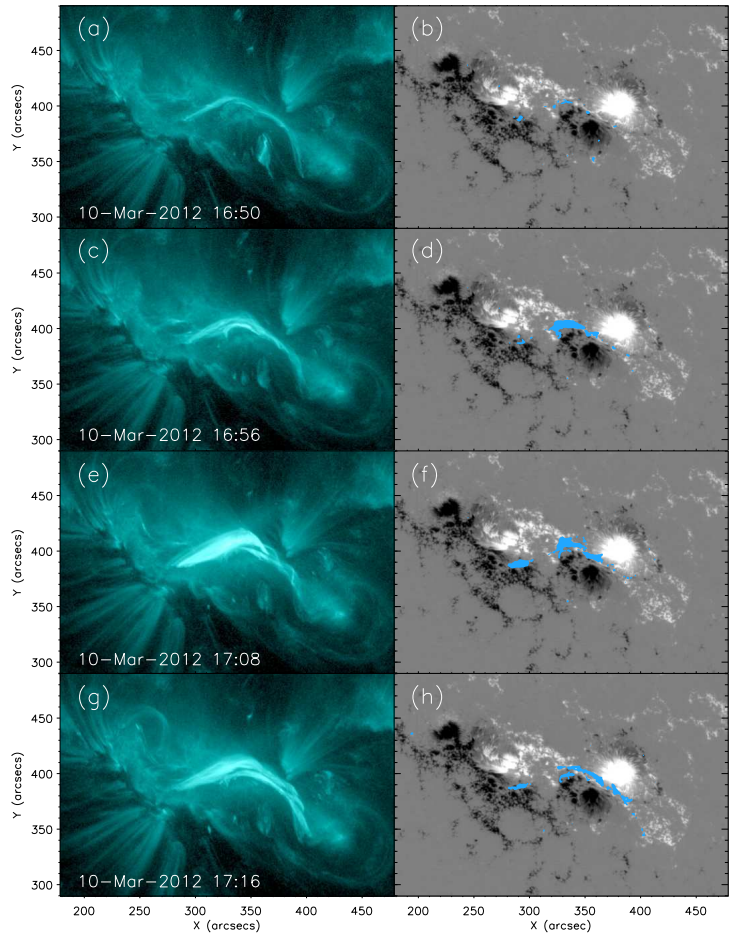}
\caption{Precursor phase of the compound eruption. The left four images show the change of the hot structure in AIA~131~\AA{} during the precursor phase. The right four panels show the flare-ribbon evolution during the pre-flare phase in 1600~\AA (blue contours), plotted over HMI/line-of-sight magnetograms (grayscale).
The sequence starts on 2012 March 10, 16:50.09 and ends on 2012 March 10, 17:16:09. The video duration is 5 s.\\
(An animation of this figure is available.)}
\label{fig:pre_flare}
\end{figure}

The precursor phase is defined as the time period during which a small amount of energy is released just prior to the IP (main energy release phase) of the solar eruption (e.g., \citealt{Zhou_etal_2016}). Generally, a small peak is observed in the time profile of the SXR flux in the precursor phase, which marks the initiation or slow rise phase of the associated CME (\citealt{Zhang_and_Dere_2006}). During the precursor phase of this event, brightenings were observed in between the two filaments, starting at around 16:33 UT. After that moment the whole HCSA became bright in 131~\AA{} (Figure~\ref{fig:pre_flare}(a); an animation accompanies Figure~\ref{fig:pre_flare}). Brightenings could have been due to the interaction between the two filaments. Evidence for such interaction was observed in a previous study of DD structures (\citealt{Liu_etal_2012}). The region near the left elbow of the HCSA was intensely brighter than the rest of the HCSA structure. In AIA~1600~\AA{} there were sporadic brightenings underneath and along the HCSA, which later concentrated more toward the left elbow around 17:08 UT (Figure \ref{fig:pre_flare}(f)). The study by \citet{Cheng_and_Ding_2016} of erupting sigmoidal structures found that the initial sporadic brightenings of the flare ribbons marked the slow rise phase of the erupting sigmoidal structure. Brightenings in 1600~\AA{} during the precursor phase may indicate the instability and the slow rise of the HCSA in the DD configuration. The slow rise may in turn indicate the initiation-phase of the solar eruption; this is observed as a small peak in the SXR time profile in this event. The intense brightening at the left elbow of the HCSA might be due to the interaction of the rising HCSA with the overlying magnetic fields. Such activities could result in a weakening of the strapping force from the overlying magnetic fields.

\subsection{\textbf{The Impulsive Phase}}
\begin{figure}[h]
\centering
\includegraphics[width=16cm,clip]{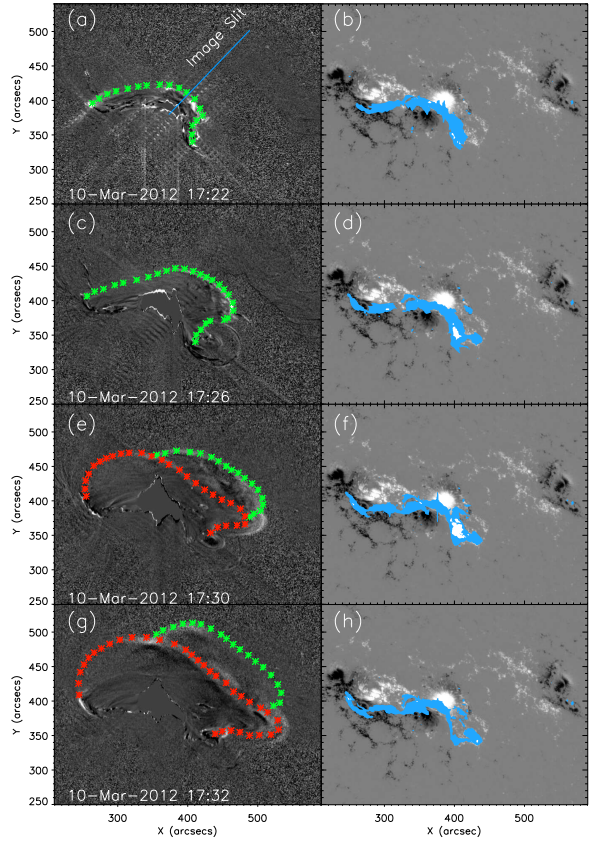}
\caption{Evolution of the compound eruption during the impulsive phase. The left four panels show the running difference images for AIA~131~\AA{} and the right four panels show the flare-ribbon evolution, overplotted on the magnetogram, in AIA~1600~\AA{} in blue. The erupting HCSA and HCSB are shown as green and red asterisks respectively. The blue line in panel (a) is the position of the slice used for the study of kinematic evolution of the two erupting structures.
The sequence starts on 2012 March 10, 17:16:21 and ends on 2012 March 10, 17:48:21 for the AIA 131 \AA\ left images. 
For the right images the flare-ribbon evolution starts on 2012 March 10, 17:16:17 
and ends on 2012 March 10, 17:48:17.  The video duration is 3 s.\\
(An animation of this figure is available.)
}
\label{fig:impulsive_phase}
\end{figure}

The IP of a solar eruption is the phase of main energy release. It is characterized by a sharp increase in SXR flux and fast acceleration of the erupting structure \citep{Zhang_and_Dere_2006}. The onset of the IP for the compound eruption of 2012 March 10 started at 17:16 UT with the acceleration of the HCSA (the latter shown with green asterisks in the AIA~131~\AA{} images in figure~\ref{fig:impulsive_phase}). The onset of the IP was also observed as an intense and continuous brightening in AIA~1600~\AA{} along the HCSA. As the IP progressed, the HCSA lifted to higher levels, accompanied by the separation of ribbons seen in 1600~\AA{} and simultaneously to the rising and writhing of filament-A (for a better view of the evolution, see the animation showing the filament eruption). This series of events accompanying the acceleration of HCSA occurred during the first IP-peak in the SXR profile (first black arrow on the left in Figure \ref{fig:three_phase}).

The HCSB began rising at around 17:23 UT. Its fast acceleration followed  at 17:29 UT as seen in AIA~131~\AA{} (shown as red asterisks in the Figure \ref{fig:impulsive_phase}). With the acceleration and rise motion of the HCSB, new sets of flare ribbons were observed which were accompanied by the rising filament-B (see the animation showing the filament eruption). The second peak in the SXR profile during the IP (second black arrow in Figure \ref{fig:three_phase}) is associated with the rise of the HCSB. The eruptions of the two hot-channel structures of the DD configuration were separated in time and thus were observed as two peaks in the IP of the SXR profile.

\begin{figure}[h]
\centering
\includegraphics[width=16cm,clip]{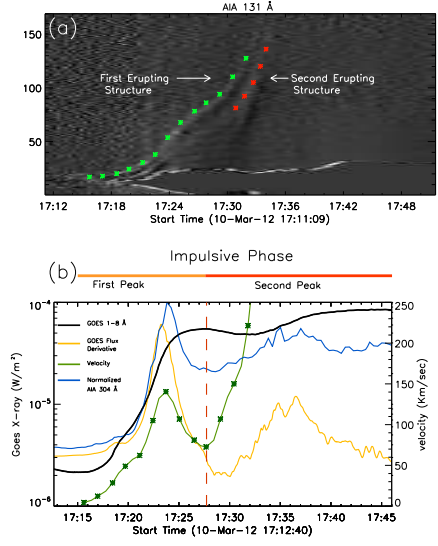}
\caption{Kinematic evolution of erupting structures. (a) The time-stacking plot obtained from the running difference of AIA~131~\AA{} images (the position of the slice is shown in Figure~\ref{fig:impulsive_phase}(a)). The green (red) asterisks mark the positions of the outer edge of the erupting HCSA (HCSB) in the DD configuration. The two erupting structures are clearly resolved. (b) The black curve is the $\textit{GOES}$ SXR profile and blue curve shows the normalized AIA~304~\AA{} intensity. The green curve shows the velocity variation of the HCSA during the impulsive phase; it has two acceleration phases. The curve in yellow is the derivative of the $\textit{GOES}$ flux. The vertical line shows the start time of the second acceleration phase of the HCSA and it corresponds to the second increase in the intensity of AIA~304~\AA.}
\label{fig:two_phase}
\end{figure}

\begin{figure}[h]
\centering
\includegraphics[width=\linewidth, clip]{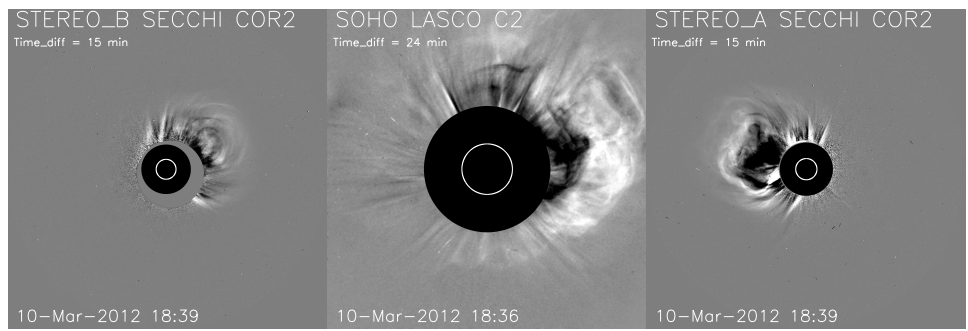}
\caption{CME resulting from the compound eruption observed from three different points of view. The three panels from left to right show COR2 image from SECCHI/STEREO-B, the C2 image from  LASCO/$\textit{SOHO}$, and the COR2 images from SECCHI/STEREO-A, respectively.}
\label{fig:corona}
\end{figure}

To quantitatively study the rising motion of the erupting structures, we chose an image slit, shown by the blue line in Figure~\ref{fig:impulsive_phase}(a), and made a time-stacking plot along this slit. In the stack-plot (see Figure~\ref{fig:two_phase}(a)) we can clearly identify the eruption of two hot-channel structures of the DD configuration. The position of the outer edge of the HCSA is shown as green asterisks in the plot in Figure~\ref{fig:two_phase}(a). The velocities obtained from these points indicate that the HCSA has undergone two acceleration phases. The first corresponds to the first peak of the IP, and the HCSA accelerated the second time when the HCSB erupted. The interaction between HCSA and HCSB seems to be responsible for the second acceleration phase of the HCSA. The HCSA had the second acceleration phase around 17:28 UT. The second acceleration phase also coincided with an increase in the normalized intensity of AIA~304~\AA{} (Figure~\ref{fig:two_phase}(b)). The velocity time profile of the HCSA and the intensity profile of the flux in AIA~304~\AA{}
shows that the fast acceleration of the HCSB started at 17:28 UT, 12 minutes after that of the HCSA. Afterwards the two structures appeared to be moving with the same velocity. The eruptions of the two structures were observed as a single CME in the $\textit{SOHO}$  LASCO/COR2 and SECCHI STEREO/COR2 coronagraph images (Figure \ref{fig:corona}).

\section{\textbf{Observations of the Pre-Eruption Evolution}}
\label{obs}
The AR 11429 was very active in terms of number and intensity of solar eruptions since it appeared on the eastern limb; it had produced 48 C-class, 15 M-class and 3 X-class flares during its transit across the solar disk. This super-activity seems to be associated with its very complex magnetic configuration, $\delta$-spot, and anti-Hale configuration (see, \citealt{Elmhamdi_etal_2014} for the evolution of the magnetic configuration). The following subsections describe the pre-eruption evolution and magnetic configuration before the onset of the compound eruption on 2012 March 10.

\subsection{\textbf{Evolution of the Photospheric Magnetic Field}}
\begin{figure}[h]
\centering
\includegraphics[width=16cm, clip]{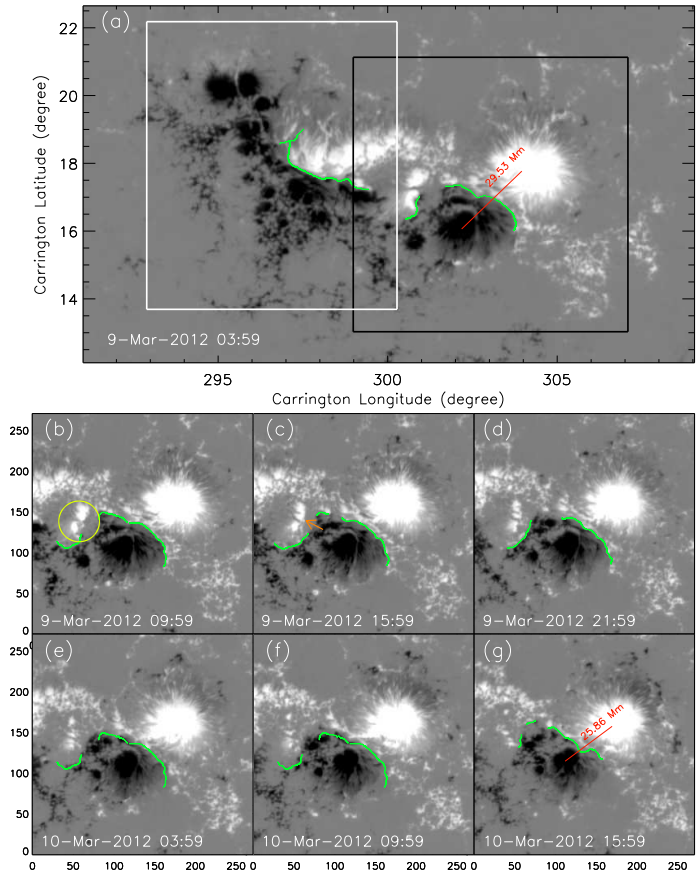}
\caption{Evolution of the photospheric magnetic field in NOAA AR 11429. Images are taken from the line-of-sight magnetograms of $\textit{SDO}$/HMI and are shown in a cylindrical equal-area heliographic projection. Green lines are the strong-gradient polarity inversion line. Panel (a) shows the entire AR, and white and black boxes show the NE and SW sub-regions respectively. Panels (b)-(g) show the evolution of the magnetic configuration in the SW part of the AR. An animation of these panels is available starting on 9 March 2012, 03:59:45 until 2012 March 10, 17:26:45. The video duration is 31 s. The red lines in panel (a) and (g), show the distance between the centroid of two strong bipoles. The yellow circle in panel (b) is the location of prominent flux cancellation. The orange arrow in panel (c) is the direction of motion of negative flux.\\
(An animation of this figure is available.)}
\label{fig:hmi_evo}
\end{figure}

To study the evolution of the magnetic configuration of AR 11429 we made use of preprocessed cutouts of the line-of-sight (LOS) HMI magnetograms. The preprocessing resolves the 180$^{\circ}$ azimuth ambiguity and remaps helioprojective images into a cylindrical equal area (CEA) projection, where each pixel has same surface area (\citealt{Hoeksema_etal_2014}). We focused on the time range between March 9, 04:00 UT to March 10, 19:00 UT. This large and complex AR could be divided into two sub-regions, NE and SW (shown as boxes in Figure~\ref{fig:hmi_evo}(a)). Both sub-regions contained a long and sharp strong-gradient PIL (shown in green lines in Figure~\ref{fig:hmi_evo}, obtained from an image gradient operation as detailed in \citealt{Zhang_etal_2010}). The compound eruption on March 10 occurred in the SW sub-region of the AR, whose evolution is shown in Figure~\ref{fig:hmi_evo}. The PIL in the SW sub-region was located in between two very strong opposite polarities. We measured the centroid of positive (negative) flux greater (smaller) than 500 G (-500G) in the SW sub-region. During the evolution of AR 11429, the distance between the centroid of the two opposite polarities (shown by red lines in Figure~\ref{fig:hmi_evo}(a) and (g)) in SW sub-region decreased from $\sim$ 29.53 Mm to $\sim$ 25.86 Mm, indicating the convergence motion of the two poles. This convergence was accompanied by fragmentation and diffusion of the magnetic flux, which resulted in further collision and cancellation of opposite flux along the PIL (see the animation accompanying Figure~\ref{fig:hmi_evo}). Flux cancellation occurred along the entire PIL; however it was most noticeable for the chunk of positive flux on the left side of the SW sub-region (shown with a yellow circle in Figure~\ref{fig:hmi_evo}(b)). The cancellation along the PIL suggests an increase of the free magnetic energy and has been associated with the formation of MFRs (\citealt{van_Ballegooijen_etal_1989, Green_etal_2011}). In addition, we observed  shearing motions along the PIL (the direction of motion is shown with an orange arrow in Figure~\ref{fig:hmi_evo}(c)), which seemed to change the semi-circular shape of the PIL to a linear-like shape. Shearing motions are also thought to increase the free magnetic energy (e.g., \citealt{Cheng_etal_2014}) in ARs. Thus, the flux cancellation and shearing motions suggest an enhancement of the free magnetic energy in the region and create the magnetic structure pending eruption.

\subsection{\textbf{Evolution of the Pre-eruption Structure in the Corona}}
\begin{figure}[h]
\centering
\includegraphics[width=\linewidth, clip]{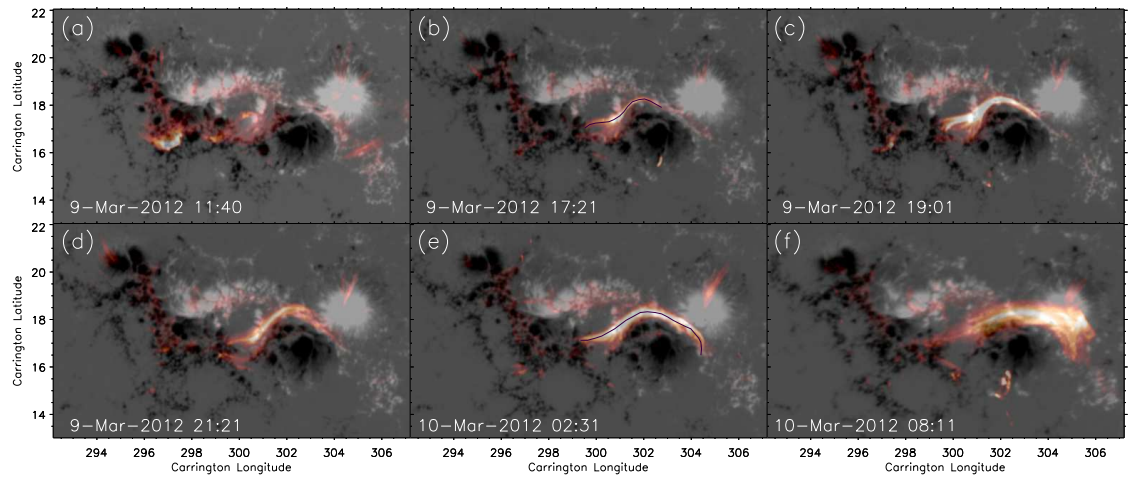}
\caption{Evolution of the coronal hot-structure before the eruption in NOAA AR 11429. AIA~131~\AA{} passband data are shown in red/white, and the HMI data are shown in white (positive flux) and black (negative flux). In panels (b) and (e) the coronal hot structures are also traced with dark lines.}
\label{fig:c_evo}
\end{figure}

During our observing period of 39 hr between March 9, 04:00 UT and March 10,  19:00 UT, many small and transient brightenings and confined flares occurred in the AR, including three confined C-class flares. This type of activity was mainly confined to the left side of the SW sub-region. There were also sporadic brightenings at the right side of the SW sub-region. Specifically, during that period (see Figure~\ref{fig:c_evo}) we observed the brightenings of coronal structures lying along the PIL. The intermittent brightenings in the AIA~131~\AA{} along the PIL indicated the presence of low-lying coherent coronal structures, which brightened up during the small confined flares. As the AR evolved, the length of the hot structure grew in size. We transformed the helioprojective images to a Carrington projection and estimated the length of hot-channel structure. The length was $\sim$ 45 Mm on 2012 March 9, at 17:21 UT and $\sim$ 76 Mm on 2012 March 10, at 02:31 UT (see Figures~\ref{fig:c_evo}(b) and (e)) as revealed by confined flares. In the corona, the length of the non-potential magnetic field lines could increase due to the tether cutting mechanism during small flare events when new connectivity could form. In the SW sub-region of the AR, converging and shearing motions may have facilitated the tether cutting reconnection and thus increased the length of the hot-channel structure. \citet{Chintzoglou_etal_2015} observed a similar kind of flare activity, small transient brightenings and confined flares, during the formation of MFRs.

\subsection{\textbf{Observations of the Double-Decker Configuration}}
\label{coronal_configuration}
\begin{figure}[tb]
\centering
\includegraphics[width=\linewidth, clip]{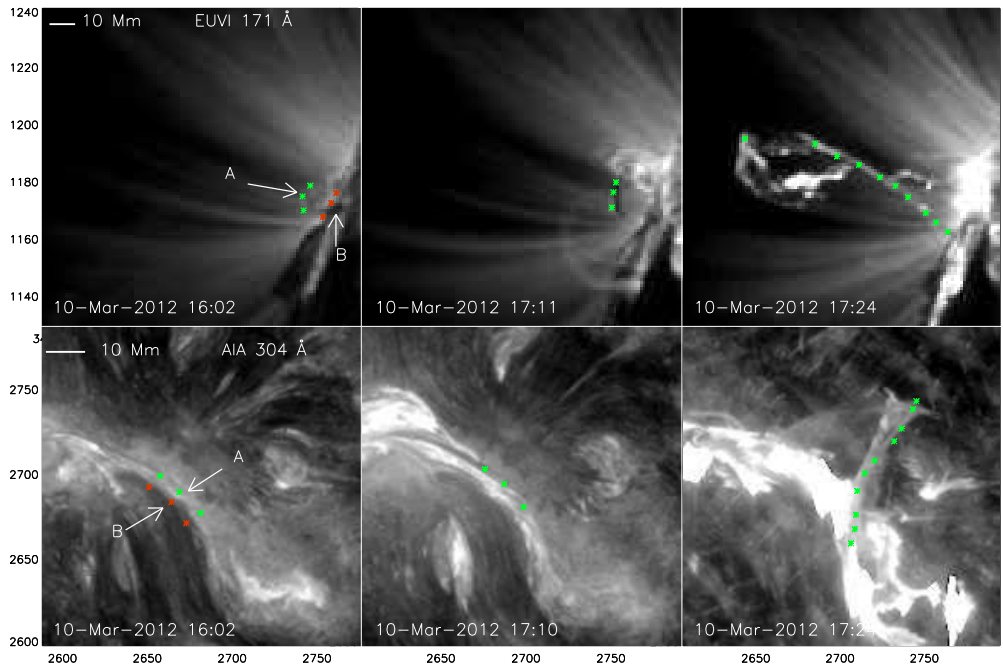}
\caption{Two erupting filaments/prominences in a DD configuration. The upper filament (A) is marked with green asterisks and the lower filament (B) with red asterisks. The upper three panels show the filaments at different time in $\textit{STEREO-A}$ EUVI~171~\AA. The lower three panels show the two filaments in $\textit{SDO}$ AIA~304~\AA. The sequence starts on 2012 March 10, 16:31:32 and ends on 2012 March 10, 18:08:32. The video duration is 31 s.\\
(An animation of this figure is available.)}
\label{fig:config}
\end{figure}

Long before the eruption on March 10, an underlying filament seemed to exist along the SW PIL. It was revealed when the hot post-flare loops, the product of an earlier eruption on March 9, at $\sim$ 03:30 UT, faded away. The existence of the filament shortly after the eruption suggests that it either did not erupt or erupted only partially. Around one hour prior to the compound eruption on March 10, we observed two filaments in the SW part of the AR, clearly seen from the $\textit{SDO}$ point of view (POV) (Figure~\ref{fig:config}). In the $\textit{SDO}$ POV, the two filaments seemed to be in very close proximity to each other; one filament (filament-A) was lying slightly off the SW PIL and the other (filament-B) was lying along it. Coronal structures above the solar disk were observed to deviate more and more from the PIL as they rose higher in the corona \citep{Cheng_and_Ding_2016}, suggesting that filament-A was probably lying higher in the corona than filament-B. To determine the geometry of the two filaments, we made use of the $\textit{STEREO-A}$ observations of them. For a specific point in one image, the $scc{\_}measure$ IDL routine (available in SolarSoftware Package; \citealt{Freeland_and_Handy_1998}) calculates the 3D LOS as observed by one vantage point (e.g. $\textit{SDO}$) and then draws the projection of this line into second image (e.g. $\textit{STEREO}$ at an angle of 109$^{\circ}$.7). The projected line from the second satellite is known as the $\emph{epipolar line}$ (see~\citealt{Inhester_2006}). Then the user can identify the location of features selected in the first image ($\textit{SDO}$) onto the second image ($\textit{STEREO}$) along this epipolar line. The 3D positions of the two filaments determined from such multiple viewpoint measurements show that they are indeed in a DD configuration (see Figure \ref{fig:config}). The separation along the vertical direction between the two filament was found to be $\sim$~12 Mm on 2012 March 10, at 16:02 UT. The study by \citet{Liu_etal_2012} observed a similar ($\sim$13 Mm) separation between the two filaments in a DD configuration.

Since there was no apparent flux emergence in the sub-region during the period of our observation, one can refute the idea of the emergence of lower magnetic structure below a higher-lying magnetic structure to form a DD configuration. Also, both parts of the DD were visible well before the initiation of the compound eruption, therefore the present case was different from the partial-eruption scenario where a single flux system divides into two parts during the eruption. It is possible that the flux cancellation and shearing motions played a role in the formation of the DD structures. Considering the initially stable configuration of the two structures, it is also possible that the axial currents in the lower and upper magnetic structures (or branches) flow in the same direction with respect to each other and the interaction between the filaments may be attractive in nature due to the $\textit{\textbf{J}}$ $\times$ $\textit{\textbf{B}}$ force as suggested by \citet{Kliem_etal_2014}. \citet{Liu_etal_2012} observed plasma being transferred from lower to the upper filament. This so-called filament-filament interaction in the DD was also reported in the study by \citet{Zhu_etal_2015}. Here, we also observed intermittent brightenings between the two filaments in AIA/$\textit{SDO}$, which may suggest a possible interaction between the two flux systems. From the POV of the $\textit{SDO}$ the separation between the filaments was so small that any plasma mass transfer could not be discerned easily. However, we speculate that there may be mass transfer between the two structures.

\begin{figure}[h]
\centering
\includegraphics[width=\linewidth,keepaspectratio]{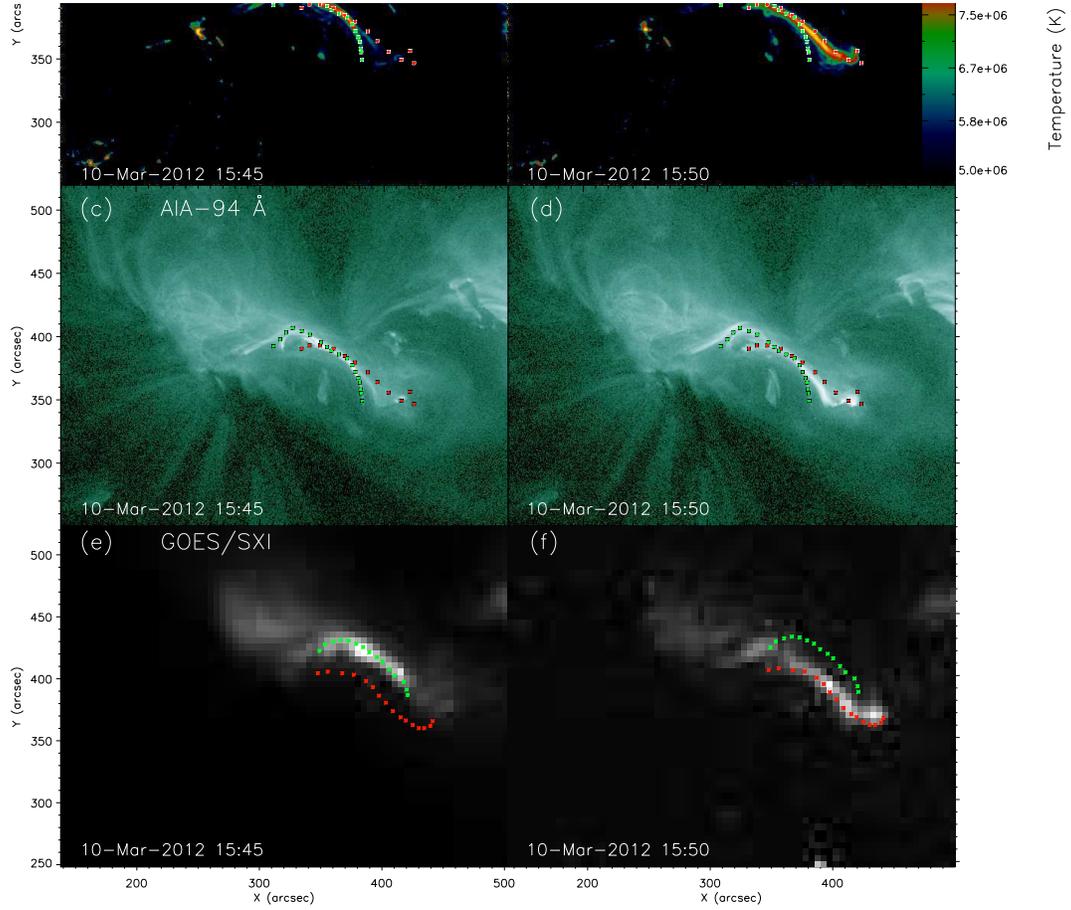}
\caption{Images illustrating the two hot-channel structures of the DD configuration in the SW region of the AR 11429. Panels (a) and (b) show the temperature map obtained using the DEM method. Panels (c) and (d) show the images in AIA~94~\AA. Panels (e) and (f) show two hot coronal structures in X-ray images from $\textit{GOES}$/SXI. The green and red asterisks show the high-lying and low-lying structures respectively.}
\label{fig:hot_config}
\end{figure}

Distinguishing the DD structures clearly in the AIA passbands against the strong on-disk background emission was a very challenging task. However, intermittent brightenings occurring at different times show two different hot-channel structures along the SW PIL (see Figure~\ref{fig:hot_config}). The temperature map, obtained using the differential emission measure (DEM) method (following \citealt{Cheng_etal_2012}; see Figure~\ref{fig:hot_config}(a) and (b)), showed that the temperature of these structures was $>$ 7 MK. Hot-channel structures were also visible in the SXR images obtained by SXI/$\textit{GOES}$ (see Figure~\ref{fig:hot_config}(e) and (f)). The foot-points of the two hot-channel structures were in close proximity to each other; however they were clearly different. Comparing the locations of the two hot-channels (AIA~94~\AA; Figures~\ref{fig:hot_config} (c) and (d)) with the two filaments (seen in AIA~304~\AA; lower panels in Figure~\ref{fig:config}), we found that the structure  marked with green asterisks (HCSA) corresponds well to filament-A and that with red asterisks (HCSB) corresponds to filament-B.

\subsection{\textbf{Coronal Magnetic Field Inferred from NLFFF Extrapolation}}
Solar eruptions are magnetic in origin so in order to better understand the associated mechanisms, it is crucial to study the magnetic configuration of the ARs (with the aid of modeling). The location of the AR 11429 on the solar disk made it a suitable candidate to study the coronal magnetic field through magnetic field extrapolation. In the solar corona the magnetic pressure is much higher than the plasma pressure and thus we can assume that the magnetic field lines satisfy the force-free field approximation (see \citealt{Wiegelmann_and_Sakurai_2012}), i.e., $\textit{\textbf{J}}$$\parallel$$\textit{\textbf{B}}$ and $\nabla$ $\times$ $\textit{\textbf{B}}$ = $\alpha$$\textit{\textbf{B}}$. Here, $\alpha$ is the force-free parameter and is constant along each magnetic field line; however it varies over different field lines in the case of NLFFFs. To get the NLFFF extrapolation of coronal magnetic field, we use an optimization numerical approach developed by \citet{Wheatland_etal_2000} and later extended by \citet{Wiegelmann_2004}. In this iterative optimization process, a ``penalty function" L = $\int\limits_{V}$ $w$(x,y,z) $\Big[$ $\textit{\textbf{B}}^{-2}$ $|$($\nabla$ $\times$ $\textit{\textbf{B}}$) $\times$ $\textit{\textbf{B}}$$|^{2}$ + $|$$\nabla$ $\cdot$ $\textit{\textbf{B}}|^{2}$$\Big]$$d^{3}$x is minimized, where $w$ is a weighting function~\citep{Wiegelmann_2004}. For the bottom boundary of the domain, we use HMI vector magnetogram cutouts. These data were preprocessed to make them suitable for the force-free condition \citep{Wiegelmann_etal_2006}. The NLFFF extrapolation was done for the entire area containing AR 11429 at every hour from 2012 March 9, 04:00 UT to 2012 March 10, 17:00 UT.

\begin{figure}[h]
\centering
\includegraphics[width=16cm, clip]{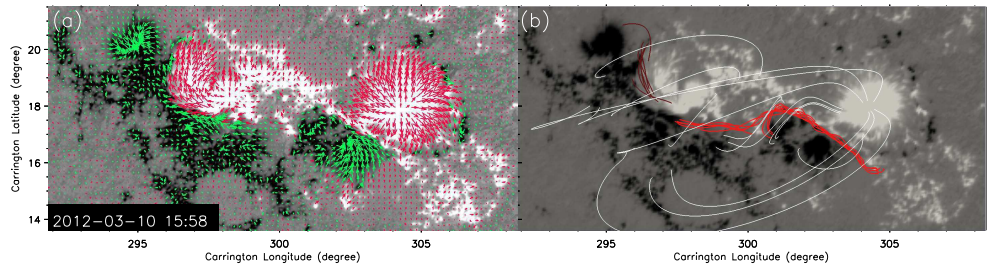}
\caption{Magnetic field configuration in the AR 11429 at 15:58 UT on 2012 March 10, immediately before the eruption. The background shows the Bz component of the vector magnetogram with positive (negative) flux in white (black). Panel (a) also shows the direction of horizontal magnetic field at the photosphere in red  and green. Panel (b) shows selected magnetic field lines in the AR above the photosphere obtained from the NLFFF extrapolation. Sheared magnetic field lines along the NE PIL are shown in brown. The sheared and twisted magnetic field lines, along the SW PIL, are shown in red. The white lines show the overlying magnetic field in the corona.}
\label{fig:mag_config}
\end{figure}

The horizontal magnetic field in the PIL was aligned along its length, indicating the presence of a strong non-potential shear along the PILs (Figure~\ref{fig:mag_config}(a)). Using the extrapolated 3D cubes and the PARAVIEW 3D visualization tool, we analyzed the evolution of the vector magnetic field in the corona. We found sheared and weakly twisted magnetic field lines above the PILs in the AR. In the NE sub-region, the sheared coronal magnetic field lines remained more or less unchanged throughout the period under study (short sheared lines shown in brown in Figure~\ref{fig:mag_config}(b)). In the SW sub-region, however, there were different groups of sheared and twisted magnetic field lines. With time, a system of long, sheared, and twisted magnetic field lines above the PIL appeared in the modeled corona data, before the compound eruption on 2012 March 10 (shown in red in Figure~\ref{fig:mag_config}(b)). The evolution of magnetic field lines was also evident from the evolution of the current density of the AR (more in the following paragraph).

\begin{figure}[tb]
\centering
\includegraphics[width=\linewidth, clip]{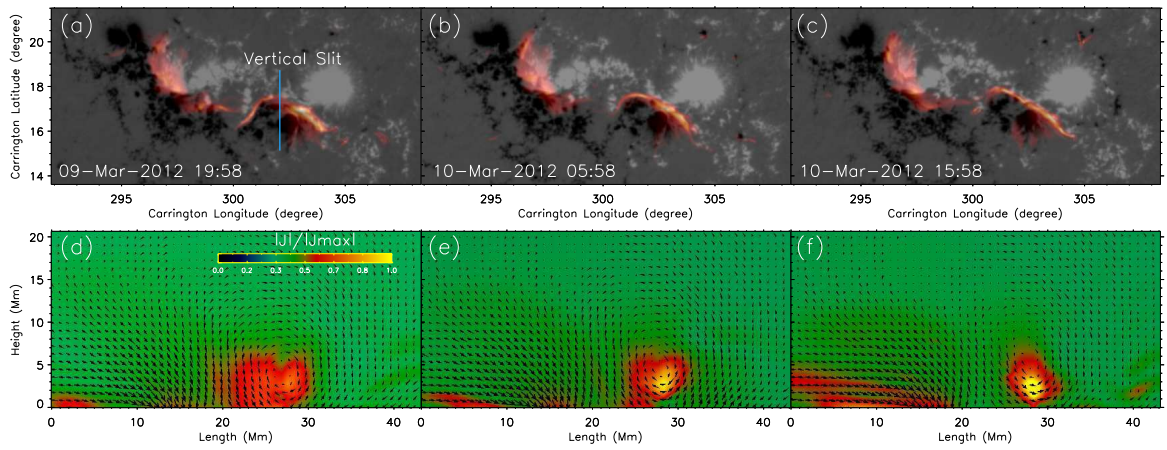}
\caption{Evolution of coronal current density (obtained from the NLFFF extrapolation) in the AR 11429. Panels (a)-(c) show the normalized-integrated current density along the line of sight in red/white and the Bz component in the background. Panels (d)-(f) show the vertical slice of the current density along the blue line (top left panel). During the evolution, the current density changed from more diffuse to more concentric around the location of the hot channel.}
\label{fig:current}
\end{figure}

We calculated the current density in the AR as $\textit{\textbf{J}}$ = $\nabla$  $\times$  $\textit{\textbf{B}}$/$\mu_o$ for the entire extrapolated domain. Along the PILs we observed very high current density. The presence of current density signifies non-potential magnetic fields. Due to Ohmic heating, regions of higher current density are hotter than the neighbouring regions. These regions are observed as diffuse and sigmoidal structures in the AIA hot-channel passbands. In the NE sub-region, the high current-density region was more diffuse and remained unchanged during the evolution of the AR. In the SW sub-region, on the other hand, the shape of the high current-density region changed significantly over time (see upper panels of Figure \ref{fig:current}, corresponding to the change in the SW PIL in Figure \ref{fig:hmi_evo}) and also became more concentrated (see the lower panels of Figure \ref{fig:current}). The change of the high current-density region from a more diffuse to more concentrated current distribution indicated the formation of long, sheared, and twisted magnetic field lines above the photosphere as displayed in the extrapolation results. This is also seen in \citet{Chintzoglou_etal_2015} where the 3D current density before the formation of the MFRs was initially a system of fragmented current channels and over time became a single current channel structure. 

The shearing motion and flux cancellation changed the magnetic configuration of the AR 11429. The extrapolation results showed that these changes were favorable for the formation of a long, sheared, and twisted magnetic structure at the SW PIL of the AR. There was one group of sheared and twisted magnetic field lines along the SW PIL, shown in red in Figure \ref{fig:mag_config}(b), corresponding to a single concentrated current density in the lower-right panel of Figure \ref{fig:current}. The DD configuration was not reproduced in the NLFFF extrapolated result. The lack of a DD configuration in the extrapolation might be due to the fact that its lower and the upper branches formed during the temporal evolution. It is also important to note that the NLFFF extrapolation is producing static equilibria from the snapshot image (coronal magnetic field for the photospheric boundary condition) of the magnetic field, thus the dynamics are not fully captured. Nevertheless, NLFFF extrapolation helps in understanding the magnetic distribution above the photosphere and slow quasi-static evolution to first order.

\section{\textbf{Discussion and Conclusion}}
\label{DC}

Our primary aim for the present study is to understand and explain the two peaks in the IP of the solar eruption from AR 11429 on 2012 March 10. It is a compound eruption, since it involves two closely spaced magnetic flux rope candidates which successively erupted within a short time interval (12 minutes in this event). The compound eruption was associated with an M-class flare. We carried out a detailed study of the compound eruption and the evolution of the AR for 39 hr leading to the eruption. Around an hour prior to the compound eruption, we observed two filaments lying along the same PIL on the SW sub-region of the AR.

The main result of the study is that the two peaks in the IP are caused by the eruption of two hot-channel structures (as seen in 131~\AA{} corresponding to 10 MK plasma emission), which were likely in the form of a flux rope during or even before the eruption. Our analysis suggests that each erupting hot-channel structure was preceded by an independent coherent magnetic structure that destabilized and erupted. Here, we call such an eruption-capable pre-eruption magnetic structure a flux bundle, which is so named to emphasize its becoming coherent and independent long before (i.e., many hours or more than one day) the eruption, but without specifying whether such a structure is a flux rope or not. Thus, both sheared structures and flux ropes can be considered as flux bundles, as long as they become an entity of eruption at a later time. These two flux bundles were lying along the same PIL one above the other in a DD configuration. The instability and fast acceleration started first for the high-lying structure, then was followed by the instability and fast acceleration of the low-lying structure. The velocity of the low-lying structure was greater than that of the high-lying structure; as a result, the second structure reached the first and both interacted with each other to form a compound eruption. The eruption of the two flux bundles and the magnetic reconnection in the current sheet underneath them was responsible for the double peaks in the IP of the associated flare.

The present case is different from classical eruptive model where a single flux-rope-like structure arises during the solar eruption and a single peak is observed in the SXR profile during the IP. A few previous studies consider the eruption of DD structures. The study by \citet{Liu_etal_2012} observed two filaments lying along the same PIL. They observed the instability and eruption of only the higher-lying filament. \citet{Zhu_etal_2015} also studied the eruption of the DD configuration. In their study the lower filament rose and merged with the upper filament and then the merged filaments erupted together. Unlike the previous studies, in this work, we observed the two components of the DD configuration that destabilized and erupted at different times and then interacted with each other to form a compound eruption. 

\begin{figure}[h]
\centering
\includegraphics[width=\linewidth,clip]{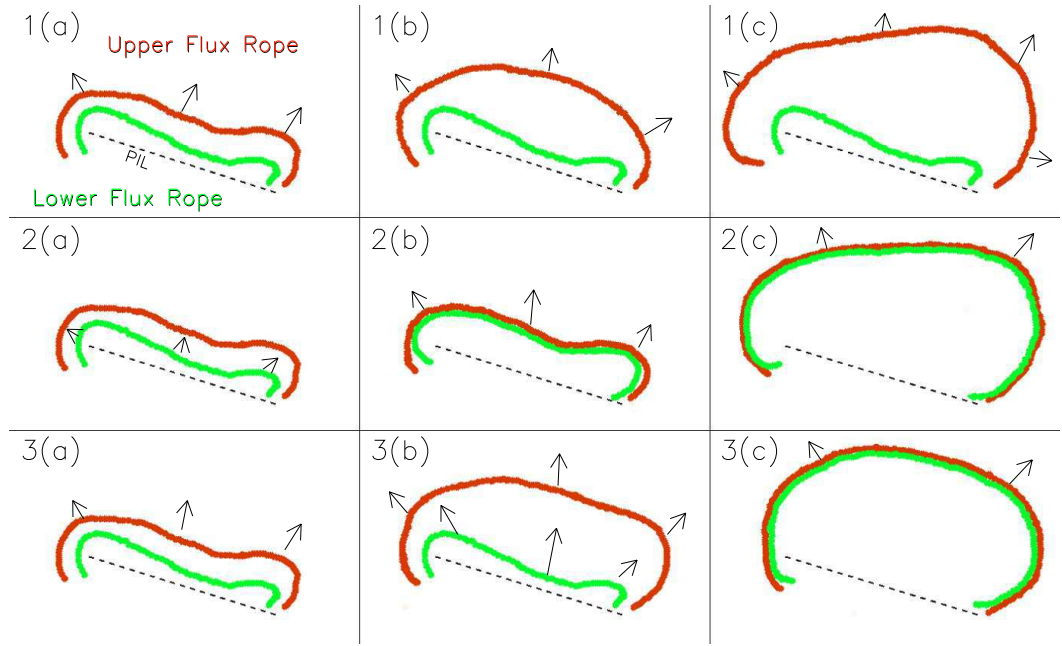}
\caption{Three erupting scenarios of the DD configuration. (1) The upper three panels depict the case where only the upper-branch of the DD erupts. (2) The middle three panels show the merging of the two branches of the DD, which erupt together. (3) The lower three panels show the eruption of the two branches of the DD separately and their interaction after the eruption.}
\label{fig:cartoon}
\end{figure}

Based on the studies mentioned above, we can summarize that there exist three different scenarios for eruptions in a DD configuration (shown in Figure \ref{fig:cartoon}). They are: (1) Instability and eruption of only the high-lying magnetic structure (this is the partial eruption of the DD configuration); (2) instability and rise of the low-lying magnetic structure, merging with the high-lying magnetic structure. Later on the merged structure erupts (this is the case of full eruption of the DD configuration); (3) instability and acceleration first of the high-lying magnetic structure followed by the low-lying magnetic structure. The two erupting structures interact with each other forming a compound structure (this is also the case of full eruption). Cases (1) and (2) would be similar to the eruption of a single flux rope with a single peak in the IP. Only case (3) would result in two peaks in the IP.

\citet{Kliem_etal_2014} studied two flux ropes in the DD configuration through MHD simulations. Analytically, they found that the DD configuration remained in stable equilibrium if the toroidal component of external sheared field lines, lying above two flux ropes, had a strength above a certain threshold value. The decrease in the strength of the toroidal component of the overlying sheared arcade would result in the instability of both of the flux ropes. Usually in such conditions the lower flux rope becomes unstable first and results in two cases: (a) merging of the two flux ropes and full eruption of the DD structure, and (b) only the upper flux rope erupts and the lower flux rope gets destroyed due to reconnection with the ambient flux. For the case when the toroidal component of the overlying sheared arcade is above the threshold value, an eruption of the upper flux rope is possible if transfer of magnetic flux and current occurs from the lower to the upper flux rope. This case also results in the eruption of the upper flux rope only (partial transfer of flux from bottom flux rope). However, in this case the lower one remains in place, possibly because it loses free energy by transferring it to the upper flux rope. In our case the interaction between the lower and the upper filament was observed as brightenings between them. During such interactions it is possible that plasma, magnetic flux and currents become transferred from the lower to the upper flux rope. Normally, such a transfer would have resulted in an eruption of only the upper flux rope, leaving the lower one undisturbed. However, the interaction of the erupting upper flux rope with the overlying sheared arcade might have reduced the toroidal component of the latter. It is possible that this interaction resulted in the instability of the lower flux rope, leading to a sequential eruption. This is one of the possible explanations of the observed eruption of the DD in the present case, but other explanations are also possible and should be addressed in a future study.

Finally, we would like to comment on the formation of the DD configuration. The sub-flaring events during the evolution of the AR possibly increased the length of coronal hot structures and formed flux bundles through magnetic reconnection. \citet{Wang_and_Zhang_2007} found that there is a low probability of having open eruption from the magnetic center of an AR due to strong overlying magnetic pressure. The SW PIL was lying between strong magnetic polarities, which implies that there was strong overlying magnetic pressure over the coronal structure above it. Due to that strong overlying magnetic pressure it was difficult for the magnetic structures to erupt. \citet{Cheng_etal_2014} observed the formation of a new flux rope below the existing one due to shearing, flux cancellation and rotation of the leading magnetic polarity. Continuous shearing and flux cancellation in the AR may have resulted in the formation of a second flux rope that was lying low in the corona. Thus, strong overlying magnetic pressure and continuous shearing and flux cancellation below a higher lying flux bundle might have been responsible for the formation of a second lower-lying flux bundle, forming the DD configuration along the same PIL. 

\acknowledgments
The authors are greatful to the anonymous referee for the careful review and valuable comments. HMI and AIA are instruments on board $\textit{SDO}$, a mission for NASA's Living With a Star Program. S.D. is supported by GMU Presidential Scholarship Award. G.C. acknowledges support by NASA contract NNG04EA00C ($\textit{SDO}$/AIA). J.Z. is supported by US NSF AGS-1249270 and NSF AGS-1156120.

\bibliography{compound_arxiv}

\begin{thebibliography}{}
\expandafter\ifx\csname natexlab\endcsname\relax\def\natexlab#1{#1}\fi
\providecommand{\url}[1]{\href{#1}{#1}}
\providecommand{\dodoi}[1]{doi:~\href{http://doi.org/#1}{\nolinkurl{#1}}}
\providecommand{\doeprint}[1]{\href{http://ascl.net/#1}{\nolinkurl{http://ascl%
.net/#1}}}
\providecommand{\doarXiv}[1]{\href{https://arxiv.org/abs/#1}{\nolinkurl{https:%
//arxiv.org/abs/#1}}}

\bibitem[{{Amari} {et~al.}(2003){Amari}, {Luciani}, {Aly}, {Mikic}, \&
  {Linker}}]{Amari_etal_I_2003}
{Amari}, T., {Luciani}, J.~F., {Aly}, J.~J., {Mikic}, Z., \& {Linker}, J. 2003,
  \apj, 585, 1073, \dodoi{10.1086/345501}

\bibitem[{{Amari} {et~al.}(2000){Amari}, {Luciani}, {Mikic}, \&
  {Linker}}]{Amari_etal_2000}
{Amari}, T., {Luciani}, J.~F., {Mikic}, Z., \& {Linker}, J. 2000, \apjl, 529,
  L49, \dodoi{10.1086/312444}

\bibitem[{{Antiochos} {et~al.}(1999){Antiochos}, {DeVore}, \&
  {Klimchuk}}]{Antiochos_etal_1999}
{Antiochos}, S.~K., {DeVore}, C.~R., \& {Klimchuk}, J.~A. 1999, \apj, 510, 485,
  \dodoi{10.1086/306563}

\bibitem[{{Aulanier} {et~al.}(2010){Aulanier}, {T{\"o}r{\"o}k}, {D{\'e}moulin},
  \& {DeLuca}}]{Aulanier_etal_2010}
{Aulanier}, G., {T{\"o}r{\"o}k}, T., {D{\'e}moulin}, P., \& {DeLuca}, E.~E.
  2010, \apj, 708, 314, \dodoi{10.1088/0004-637X/708/1/314}

\bibitem[{{Biesecker} \& {Thompson}(2000)}]{Biesecker_and_Thompson_2000}
{Biesecker}, D.~A., \& {Thompson}, B.~J. 2000, Journal of Atmospheric and
  Solar-Terrestrial Physics, 62, 1449, \dodoi{10.1016/S1364-6826(00)00085-7}

\bibitem[{{Brueckner} {et~al.}(1995){Brueckner}, {Howard}, {Koomen},
  {Korendyke}, {Michels}, {Moses}, {Socker}, {Dere}, {Lamy}, {Llebaria},
  {Bout}, {Schwenn}, {Simnett}, {Bedford}, \& {Eyles}}]{Brueckner_etal_1995}
{Brueckner}, G.~E., {Howard}, R.~A., {Koomen}, M.~J., {et~al.} 1995, \solphys,
  162, 357, \dodoi{10.1007/BF00733434}

\bibitem[{{Chen}(1989)}]{Chen_1989}
{Chen}, J. 1989, \apj, 338, 453, \dodoi{10.1086/167211}

\bibitem[{{Chen} {et~al.}(1997){Chen}, {Howard}, {Brueckner}, {Santoro},
  {Krall}, {Paswaters}, {St.~Cyr}, {Schwenn}, {Lamy}, \&
  {Simnett}}]{Chen_etal_1997}
{Chen}, J., {Howard}, R.~A., {Brueckner}, G.~E., {et~al.} 1997, \apjl, 490,
  L191, \dodoi{10.1086/311029}

\bibitem[{{Cheng} \& {Ding}(2016)}]{Cheng_and_Ding_2016}
{Cheng}, X., \& {Ding}, M.~D. 2016, \apjs, 225, 16,
  \dodoi{10.3847/0067-0049/225/1/16}

\bibitem[{{Cheng} {et~al.}(2014){Cheng}, {Ding}, {Zhang}, {Sun}, {Guo}, {Wang},
  {Kliem}, \& {Deng}}]{Cheng_etal_2014}
{Cheng}, X., {Ding}, M.~D., {Zhang}, J., {et~al.} 2014, \apj, 789, 93,
  \dodoi{10.1088/0004-637X/789/2/93}

\bibitem[{{Cheng} {et~al.}(2011){Cheng}, {Zhang}, {Liu}, \&
  {Ding}}]{Cheng_etal_2011}
{Cheng}, X., {Zhang}, J., {Liu}, Y., \& {Ding}, M.~D. 2011, \apjl, 732, L25,
  \dodoi{10.1088/2041-8205/732/2/L25}

\bibitem[{{Cheng} {et~al.}(2012){Cheng}, {Zhang}, {Saar}, \&
  {Ding}}]{Cheng_etal_2012}
{Cheng}, X., {Zhang}, J., {Saar}, S.~H., \& {Ding}, M.~D. 2012, \apj, 761, 62,
  \dodoi{10.1088/0004-637X/761/1/62}

\bibitem[{{Chintzoglou} {et~al.}(2015){Chintzoglou}, {Patsourakos}, \&
  {Vourlidas}}]{Chintzoglou_etal_2015}
{Chintzoglou}, G., {Patsourakos}, S., \& {Vourlidas}, A. 2015, \apj, 809, 34,
  \dodoi{10.1088/0004-637X/809/1/34}

\bibitem[{{Elmhamdi} {et~al.}(2014){Elmhamdi}, {Romano}, {Kordi}, \&
  {Al-trabulsy}}]{Elmhamdi_etal_2014}
{Elmhamdi}, A., {Romano}, P., {Kordi}, A.~S., \& {Al-trabulsy}, H.~A. 2014,
  \solphys, 289, 2957, \dodoi{10.1007/s11207-014-0507-9}

\bibitem[{{Fan} \& {Gibson}(2003)}]{Fan_and_Gibson_2003}
{Fan}, Y., \& {Gibson}, S.~E. 2003, \apjl, 589, L105, \dodoi{10.1086/375834}

\bibitem[{{Fletcher} {et~al.}(2011){Fletcher}, {Dennis}, {Hudson}, {Krucker},
  {Phillips}, {Veronig}, {Battaglia}, {Bone}, {Caspi}, {Chen}, {Gallagher},
  {Grigis}, {Ji}, {Liu}, {Milligan}, \& {Temmer}}]{Fletcher_etal_2011}
{Fletcher}, L., {Dennis}, B.~R., {Hudson}, H.~S., {et~al.} 2011, \ssr, 159, 19,
  \dodoi{10.1007/s11214-010-9701-8}

\bibitem[{{Forbes}(2000)}]{Forbes_etal_2000}
{Forbes}, T.~G. 2000, \jgr, 105, 23153, \dodoi{10.1029/2000JA000005}

\bibitem[{{Freeland} \& {Handy}(1998)}]{Freeland_and_Handy_1998}
{Freeland}, S.~L., \& {Handy}, B.~N. 1998, \solphys, 182, 497,
  \dodoi{10.1023/A:1005038224881}

\bibitem[{{Gibson} \& {Fan}(2006)}]{Gibson_and_Fan_2006}
{Gibson}, S.~E., \& {Fan}, Y. 2006, \apjl, 637, L65, \dodoi{10.1086/500452}

\bibitem[{{Gibson} {et~al.}(2002){Gibson}, {Fletcher}, {Del Zanna}, {Pike},
  {Mason}, {Mandrini}, {D{\'e}moulin}, {Gilbert}, {Burkepile}, {Holzer},
  {Alexander}, {Liu}, {Nitta}, {Qiu}, {Schmieder}, \&
  {Thompson}}]{Gibson_etal_2002}
{Gibson}, S.~E., {Fletcher}, L., {Del Zanna}, G., {et~al.} 2002, \apj, 574,
  1021, \dodoi{10.1086/341090}

\bibitem[{{Gilbert} {et~al.}(2000){Gilbert}, {Holzer}, {Burkepile}, \&
  {Hundhausen}}]{Gilbert_etal_2000}
{Gilbert}, H.~R., {Holzer}, T.~E., {Burkepile}, J.~T., \& {Hundhausen}, A.~J.
  2000, \apj, 537, 503, \dodoi{10.1086/309030}

\bibitem[{{Gopalswamy} {et~al.}(2002){Gopalswamy}, {Yashiro}, {Micha{\l}ek},
  {Kaiser}, {Howard}, {Reames}, {Leske}, \& {von
  Rosenvinge}}]{Gopalswamy_etal_2002}
{Gopalswamy}, N., {Yashiro}, S., {Micha{\l}ek}, G., {et~al.} 2002, \apjl, 572,
  L103, \dodoi{10.1086/341601}

\bibitem[{{Green} {et~al.}(2007){Green}, {Kliem}, {T{\"o}r{\"o}k}, {van
  Driel-Gesztelyi}, \& {Attrill}}]{Green_etal_2007}
{Green}, L.~M., {Kliem}, B., {T{\"o}r{\"o}k}, T., {van Driel-Gesztelyi}, L., \&
  {Attrill}, G.~D.~R. 2007, \solphys, 246, 365,
  \dodoi{10.1007/s11207-007-9061-z}

\bibitem[{{Green} {et~al.}(2011){Green}, {Kliem}, \&
  {Wallace}}]{Green_etal_2011}
{Green}, L.~M., {Kliem}, B., \& {Wallace}, A.~J. 2011, \aap, 526, A2,
  \dodoi{10.1051/0004-6361/201015146}

\bibitem[{{Hill} {et~al.}(2005){Hill}, {Pizzo}, {Balch}, {Biesecker},
  {Bornmann}, {Hildner}, {Lewis}, {Grubb}, {Husler}, {Prendergast}, {Vickroy},
  {Greer}, {Defoor}, {Wilkinson}, {Hooker}, {Mulligan}, {Chipman}, {Bysal},
  {Douglas}, {Reynolds}, {Davis}, {Wallace}, {Russell}, {Freestone},
  {Bagdigian}, {Page}, {Kerns}, {Hoffman}, {Cauffman}, {Davis}, {Studer},
  {Berthiaume}, {Saha}, {Berthiume}, {Farthing}, \&
  {Zimmermann}}]{Hill_etal_2005}
{Hill}, S.~M., {Pizzo}, V.~J., {Balch}, C.~C., {et~al.} 2005, \solphys, 226,
  255, \dodoi{10.1007/s11207-005-7416-x}

\bibitem[{{Hoeksema} {et~al.}(2014){Hoeksema}, {Liu}, {Hayashi}, {Sun},
  {Schou}, {Couvidat}, {Norton}, {Bobra}, {Centeno}, {Leka}, {Barnes}, \&
  {Turmon}}]{Hoeksema_etal_2014}
{Hoeksema}, J.~T., {Liu}, Y., {Hayashi}, K., {et~al.} 2014, \solphys, 289,
  3483, \dodoi{10.1007/s11207-014-0516-8}

\bibitem[{{Howard} {et~al.}(2008){Howard}, {Moses}, {Vourlidas}, {Newmark},
  {Socker}, {Plunkett}, {Korendyke}, {Cook}, {Hurley}, {Davila}, {Thompson},
  {St Cyr}, {Mentzell}, {Mehalick}, {Lemen}, {Wuelser}, {Duncan}, {Tarbell},
  {Wolfson}, {Moore}, {Harrison}, {Waltham}, {Lang}, {Davis}, {Eyles},
  {Mapson-Menard}, {Simnett}, {Halain}, {Defise}, {Mazy}, {Rochus}, {Mercier},
  {Ravet}, {Delmotte}, {Auchere}, {Delaboudiniere}, {Bothmer}, {Deutsch},
  {Wang}, {Rich}, {Cooper}, {Stephens}, {Maahs}, {Baugh}, {McMullin}, \&
  {Carter}}]{Howard_etal_2008}
{Howard}, R.~A., {Moses}, J.~D., {Vourlidas}, A., {et~al.} 2008, \ssr, 136, 67,
  \dodoi{10.1007/s11214-008-9341-4}

\bibitem[{{Inhester}(2006)}]{Inhester_2006}
{Inhester}, B. 2006, ArXiv Astrophysics e-prints

\bibitem[{{Jacobs} {et~al.}(2009){Jacobs}, {Roussev}, {Lugaz}, \&
  {Poedts}}]{Jacobs_etal_2009}
{Jacobs}, C., {Roussev}, I.~I., {Lugaz}, N., \& {Poedts}, S. 2009, \apjl, 695,
  L171, \dodoi{10.1088/0004-637X/695/2/L171}

\bibitem[{{Kaiser} {et~al.}(2008){Kaiser}, {Kucera}, {Davila}, {St.~Cyr},
  {Guhathakurta}, \& {Christian}}]{Kaiser_etal_2008}
{Kaiser}, M.~L., {Kucera}, T.~A., {Davila}, J.~M., {et~al.} 2008, \ssr, 136, 5,
  \dodoi{10.1007/s11214-007-9277-0}

\bibitem[{{Kliem} {et~al.}(2014){Kliem}, {T{\"o}r{\"o}k}, {Titov}, {Lionello},
  {Linker}, {Liu}, {Liu}, \& {Wang}}]{Kliem_etal_2014}
{Kliem}, B., {T{\"o}r{\"o}k}, T., {Titov}, V.~S., {et~al.} 2014, \apj, 792,
  107, \dodoi{10.1088/0004-637X/792/2/107}

\bibitem[{{Leake} {et~al.}(2014){Leake}, {Linton}, \&
  {Antiochos}}]{Leake_etal_II_2014}
{Leake}, J.~E., {Linton}, M.~G., \& {Antiochos}, S.~K. 2014, \apj, 787, 46,
  \dodoi{10.1088/0004-637X/787/1/46}

\bibitem[{{Leake} {et~al.}(2013){Leake}, {Linton}, \&
  {T{\"o}r{\"o}k}}]{Leake_etal_I_2013}
{Leake}, J.~E., {Linton}, M.~G., \& {T{\"o}r{\"o}k}, T. 2013, \apj, 778, 99,
  \dodoi{10.1088/0004-637X/778/2/99}

\bibitem[{{Lemen} {et~al.}(2012){Lemen}, {Title}, {Akin}, {Boerner}, {Chou},
  {Drake}, {Duncan}, {Edwards}, {Friedlaender}, {Heyman}, {Hurlburt}, {Katz},
  {Kushner}, {Levay}, {Lindgren}, {Mathur}, {McFeaters}, {Mitchell}, {Rehse},
  {Schrijver}, {Springer}, {Stern}, {Tarbell}, {Wuelser}, {Wolfson}, {Yanari},
  {Bookbinder}, {Cheimets}, {Caldwell}, {Deluca}, {Gates}, {Golub}, {Park},
  {Podgorski}, {Bush}, {Scherrer}, {Gummin}, {Smith}, {Auker}, {Jerram},
  {Pool}, {Soufli}, {Windt}, {Beardsley}, {Clapp}, {Lang}, \&
  {Waltham}}]{Lemen_etal_2012}
{Lemen}, J.~R., {Title}, A.~M., {Akin}, D.~J., {et~al.} 2012, \solphys, 275,
  17, \dodoi{10.1007/s11207-011-9776-8}

\bibitem[{{Lites}(2005)}]{Lites_etal_2005}
{Lites}, B.~W. 2005, \apj, 622, 1275, \dodoi{10.1086/428080}

\bibitem[{{Liu} {et~al.}(2012){Liu}, {Kliem}, {T{\"o}r{\"o}k}, {Liu}, {Titov},
  {Lionello}, {Linker}, \& {Wang}}]{Liu_etal_2012}
{Liu}, R., {Kliem}, B., {T{\"o}r{\"o}k}, T., {et~al.} 2012, \apj, 756, 59,
  \dodoi{10.1088/0004-637X/756/1/59}

\bibitem[{{Liu} {et~al.}(2009){Liu}, {Su}, {Xu}, {Lin}, {Shibata}, \&
  {Kurokawa}}]{Liu_etal_2009}
{Liu}, Y., {Su}, J., {Xu}, Z., {et~al.} 2009, \apjl, 696, L70,
  \dodoi{10.1088/0004-637X/696/1/L70}

\bibitem[{{L{\'o}pez Ariste} {et~al.}(2006){L{\'o}pez Ariste}, {Aulanier},
  {Schmieder}, \& {Sainz Dalda}}]{Lopez_Ariste_etal_2006}
{L{\'o}pez Ariste}, A., {Aulanier}, G., {Schmieder}, B., \& {Sainz Dalda}, A.
  2006, \aap, 456, 725, \dodoi{10.1051/0004-6361:20064923}

\bibitem[{{Lynch} {et~al.}(2008){Lynch}, {Antiochos}, {DeVore}, {Luhmann}, \&
  {Zurbuchen}}]{Lynch_etal_2008}
{Lynch}, B.~J., {Antiochos}, S.~K., {DeVore}, C.~R., {Luhmann}, J.~G., \&
  {Zurbuchen}, T.~H. 2008, \apj, 683, 1192, \dodoi{10.1086/589738}

\bibitem[{{Nitta} \& {Hudson}(2001)}]{Nitta_and_Hudson_2001}
{Nitta}, N.~V., \& {Hudson}, H.~S. 2001, \grl, 28, 3801,
  \dodoi{10.1029/2001GL013261}

\bibitem[{{Patsourakos} {et~al.}(2013){Patsourakos}, {Vourlidas}, \&
  {Stenborg}}]{Patsourakos_etal_2013}
{Patsourakos}, S., {Vourlidas}, A., \& {Stenborg}, G. 2013, \apj, 764, 125,
  \dodoi{10.1088/0004-637X/764/2/125}

\bibitem[{{Pesnell} {et~al.}(2012){Pesnell}, {Thompson}, \&
  {Chamberlin}}]{Pesnell_etal_2012}
{Pesnell}, W.~D., {Thompson}, B.~J., \& {Chamberlin}, P.~C. 2012, \solphys,
  275, 3, \dodoi{10.1007/s11207-011-9841-3}

\bibitem[{{Pizzo} {et~al.}(2005){Pizzo}, {Hill}, {Balch}, {Biesecker},
  {Bornmann}, {Hildner}, {Grubb}, {Chipman}, {Davis}, {Wallace}, {Russell},
  {Cauffman}, {Saha}, \& {Berthiume}}]{Pizzo_etal_2005}
{Pizzo}, V.~J., {Hill}, S.~M., {Balch}, C.~C., {et~al.} 2005, \solphys, 226,
  283, \dodoi{10.1007/s11207-005-7417-9}

\bibitem[{{Rust} \& {Kumar}(1996)}]{Rust_and_Kumar_1996}
{Rust}, D.~M., \& {Kumar}, A. 1996, \apjl, 464, L199, \dodoi{10.1086/310118}

\bibitem[{{Schou} {et~al.}(2012){Schou}, {Scherrer}, {Bush}, {Wachter},
  {Couvidat}, {Rabello-Soares}, {Bogart}, {Hoeksema}, {Liu}, {Duvall}, {Akin},
  {Allard}, {Miles}, {Rairden}, {Shine}, {Tarbell}, {Title}, {Wolfson},
  {Elmore}, {Norton}, \& {Tomczyk}}]{Schou_etal_2012}
{Schou}, J., {Scherrer}, P.~H., {Bush}, R.~I., {et~al.} 2012, \solphys, 275,
  229, \dodoi{10.1007/s11207-011-9842-2}

\bibitem[{Shen {et~al.}(2011)Shen, Liu, \& Liu}]{Shen_etal_2011}
Shen, Y.-D., Liu, Y., \& Liu, R. 2011, Research in Astronomy and Astrophysics,
  11, 594

\bibitem[{{Song} {et~al.}(2014){Song}, {Zhang}, {Cheng}, {Chen}, {Liu}, {Wang},
  \& {Li}}]{Song_etal_2014}
{Song}, H.~Q., {Zhang}, J., {Cheng}, X., {et~al.} 2014, \apj, 784, 48,
  \dodoi{10.1088/0004-637X/784/1/48}

\bibitem[{{Sun} {et~al.}(2015){Sun}, {Bobra}, {Hoeksema}, {Liu}, {Li}, {Shen},
  {Couvidat}, {Norton}, \& {Fisher}}]{Sun_etal_2015}
{Sun}, X., {Bobra}, M.~G., {Hoeksema}, J.~T., {et~al.} 2015, \apjl, 804, L28,
  \dodoi{10.1088/2041-8205/804/2/L28}

\bibitem[{{Tang}(1986)}]{Tang_1986}
{Tang}, F. 1986, \solphys, 105, 399, \dodoi{10.1007/BF00172056}

\bibitem[{{T{\"o}r{\"o}k} {et~al.}(2011){T{\"o}r{\"o}k}, {Panasenco}, {Titov},
  {Miki{\'c}}, {Reeves}, {Velli}, {Linker}, \& {De Toma}}]{Torok_etal_2011}
{T{\"o}r{\"o}k}, T., {Panasenco}, O., {Titov}, V.~S., {et~al.} 2011, \apjl,
  739, L63, \dodoi{10.1088/2041-8205/739/2/L63}

\bibitem[{{van Ballegooijen} \& {Martens}(1989)}]{van_Ballegooijen_etal_1989}
{van Ballegooijen}, A.~A., \& {Martens}, P.~C.~H. 1989, \apj, 343, 971,
  \dodoi{10.1086/167766}

\bibitem[{{Vemareddy}(2017)}]{Vemareddy_2017}
{Vemareddy}, P. 2017, \apj, 845, 59, \dodoi{10.3847/1538-4357/aa7ff4}

\bibitem[{{Wang} \& {Zhang}(2007)}]{Wang_and_Zhang_2007}
{Wang}, Y., \& {Zhang}, J. 2007, \apj, 665, 1428, \dodoi{10.1086/519765}

\bibitem[{{Wheatland} {et~al.}(2000){Wheatland}, {Sturrock}, \&
  {Roumeliotis}}]{Wheatland_etal_2000}
{Wheatland}, M.~S., {Sturrock}, P.~A., \& {Roumeliotis}, G. 2000, \apj, 540,
  1150, \dodoi{10.1086/309355}

\bibitem[{{Wiegelmann}(2004)}]{Wiegelmann_2004}
{Wiegelmann}, T. 2004, \solphys, 219, 87,
  \dodoi{10.1023/B:SOLA.0000021799.39465.36}

\bibitem[{{Wiegelmann} {et~al.}(2006){Wiegelmann}, {Inhester}, \&
  {Sakurai}}]{Wiegelmann_etal_2006}
{Wiegelmann}, T., {Inhester}, B., \& {Sakurai}, T. 2006, \solphys, 233, 215,
  \dodoi{10.1007/s11207-006-2092-z}

\bibitem[{{Wiegelmann} \& {Sakurai}(2012)}]{Wiegelmann_and_Sakurai_2012}
{Wiegelmann}, T., \& {Sakurai}, T. 2012, Living Reviews in Solar Physics, 9, 5,
  \dodoi{10.12942/lrsp-2012-5}

\bibitem[{{Wuelser} {et~al.}(2004){Wuelser}, {Lemen}, {Tarbell}, {Wolfson},
  {Cannon}, {Carpenter}, {Duncan}, {Gradwohl}, {Meyer}, {Moore}, {Navarro},
  {Pearson}, {Rossi}, {Springer}, {Howard}, {Moses}, {Newmark},
  {Delaboudiniere}, {Artzner}, {Auchere}, {Bougnet}, {Bouyries}, {Bridou},
  {Clotaire}, {Colas}, {Delmotte}, {Jerome}, {Lamare}, {Mercier}, {Mullot},
  {Ravet}, {Song}, {Bothmer}, \& {Deutsch}}]{Wuelser_etal_2004}
{Wuelser}, J.-P., {Lemen}, J.~R., {Tarbell}, T.~D., {et~al.} 2004, in
  \procspie, Vol. 5171, Telescopes and Instrumentation for Solar Astrophysics,
  ed. S.~{Fineschi} \& M.~A. {Gummin}, 111--122

\bibitem[{{Zhang} {et~al.}(2012){Zhang}, {Cheng}, \& {Ding}}]{Zhang_etal_2012}
{Zhang}, J., {Cheng}, X., \& {Ding}, M.-D. 2012, Nature Communications, 3, 747,
  \dodoi{10.1038/ncomms1753}

\bibitem[{{Zhang} \& {Dere}(2006)}]{Zhang_and_Dere_2006}
{Zhang}, J., \& {Dere}, K.~P. 2006, \apj, 649, 1100, \dodoi{10.1086/506903}

\bibitem[{{Zhang} {et~al.}(2001){Zhang}, {Dere}, {Howard}, {Kundu}, \&
  {White}}]{Zhang_etal_2001}
{Zhang}, J., {Dere}, K.~P., {Howard}, R.~A., {Kundu}, M.~R., \& {White}, S.~M.
  2001, \apj, 559, 452, \dodoi{10.1086/322405}

\bibitem[{{Zhang} \& {Wang}(2002)}]{Zhang_and_Wang_2002}
{Zhang}, J., \& {Wang}, J. 2002, \apjl, 566, L117, \dodoi{10.1086/339660}

\bibitem[{{Zhang} {et~al.}(2010){Zhang}, {Wang}, \& {Liu}}]{Zhang_etal_2010}
{Zhang}, J., {Wang}, Y., \& {Liu}, Y. 2010, \apj, 723, 1006,
  \dodoi{10.1088/0004-637X/723/2/1006}

\bibitem[{{Zhou} {et~al.}(2016){Zhou}, {Zhang}, \& {Wang}}]{Zhou_etal_2016}
{Zhou}, G.~P., {Zhang}, J., \& {Wang}, J.~X. 2016, \apjl, 823, L19,
  \dodoi{10.3847/2041-8205/823/1/L19}

\bibitem[{{Zhu} {et~al.}(2015){Zhu}, {Liu}, {Alexander}, {Sun}, \&
  {McAteer}}]{Zhu_etal_2015}
{Zhu}, C., {Liu}, R., {Alexander}, D., {Sun}, X., \& {McAteer}, R.~T.~J. 2015,
  \apj, 813, 60, \dodoi{10.1088/0004-637X/813/1/60}

\end{thebibliography}
\bibliographystyle{aasjournal}



\end{document}